\documentstyle[psfig,11pt]{article}

\newcommand{\ble}{\begin{equation}}
\newcommand{\ele}{\end{equation}}
\newcommand{\bee}{\begin{displaymath}}
\newcommand{\eee}{\end{displaymath}}

\newcommand{\mcomma}{\mbox{\ \ ,}}
\newcommand{\pfig}[1]{\centerline{\psfig{#1}}}
\newcommand{\laur}{\marginpar{\raisebox{3in}{\small \tt LAUR 92-3053}}}

\title {
Phase shift in experimental trajectory scaling functions}
\author {Ronnie Mainieri\thanks{\tt ronnie@goshawk.lanl.gov}\  and
Robert E. Ecke\thanks{\tt ecke@goshawk.lanl.gov} \\
Center for Nonlinear Studies, MS B258, \\
Los Alamos National Laboratory, Los Alamos, NM 87545 \\
}

\begin{document}

\maketitle
\laur

\begin{abstract}
For one dimensional maps the trajectory scaling functions is invariant
under coordinate transformations and can be used to compute any 
ergodic average.  It is the most stringent test between theory and
experiment, but so far it has proven difficult to extract from
experimental data.  It is shown that the main difficulty is a
dephasing of the experimental orbit which can be corrected by
reconstructing the dynamics from several time series.  From the
reconstructed dynamics the scaling function can be accurately
extracted.
\end{abstract}

In non-linear dynamical systems it is often necessary to compare a
complicated geometrical object (a fractal) obtained from an experiment
with a similar object predicted by theory.  In general the comparison
must be made up to arbitrary coordinate transformations
(diffeomorphisms) due to the phase space reconstruction technique
\cite{FarmerReconstruction,TakensReconstruction} which does not
preserve geometrical shape.   The accepted technique to compare two
multifractals is to compare their spectra of scalings --- the
$f(\alpha)$ curves.  The advantage of using the spectrum of scalings is
that is simple to apply, fairly robust to noise, and permits one to
determine if two multifractal are {\em not} smooth deformations of each
other.  If the spectra of the experiment and theory disagree, then the
two sets cannot be diffeomorphic.  But if the two spectra agree nothing
can be said about the equality of the multifractals, as there are many
different sets which are not diffeomorphic and have the same
$f(\alpha)$ spectrum of scalings \cite{FeigenbaumStrangeSets}.

The appropriate test for the equality of multifractals obtained from
dynamical systems is to compare the invariants that characterize the
orbits that generates the multifractals.  This would be equivalent to
comparing the frequencies between two (one-dimensional) harmonic oscillators
that differ by a coordinate transformation; if the frequencies are the
same, then the orbits are equivalent.  For general dynamical systems
not all the invariants are known and the comparison is difficult.  For
hyperbolic systems, it has been conjectured
\cite{CvitanovicCycles} that the periodic orbits constitute a complete
invariant characterization.  For the period
doubling case and the circle map case it has been shown that the
scaling function introduced by Feigenbaum \cite{FeigenbaumScaling} is
the complete invariant information about the dynamical system (see
Sullivan in \cite{SullivanMaximal}).  The scaling function $\sigma(\tau)$
gives the local contraction after traversing a fraction $\tau$ of a
asymptotically long periodic orbit.  Using the scaling function and an
initial portion of an orbit, its asymptotic behavior can be
computed, and from it any phase space average.

An analogy with statistical mechanics may help elucidate the role of
the scaling function.  In the analogy the role of the circle map is
played by a one dimensional Ising model with long range (exponentially
decaying) interactions.  In this analogy the scaling function is the
interaction between the spins.  To determine the free energy of the
Ising model at a given temperature, one constructs the transfer matrix
$T$ from the interaction and determines its largest eigenvalue.  Its
logarithm is the free energy.  In the dynamical system case a matrix
can be built from the scaling function and its largest eigenvalue
determines the thermodynamics ($f(\alpha)$ spectrum, fractal
dimensions, generalized entropies).  There is more to this analogy than
the mere similarity of concepts, and by pursuing it one realizes the
central role played by the scaling function.  More details can be found
in the articles by Vul {\em et al.}~\cite{SinaiReview} and
Feigenbaum~\cite{FeigenbaumStrangeSets}, and in the book of
Ruelle~\cite{Thermodynamic}.

To compare an experiment on period doubling or on golden mean
mode locking one should extract the scaling function from the
experimental data and compare it with the theoretically computed
scaling function.  In spite of the large number of experiments
on period doubling and on mode locking \cite{EckeComo}, this has
not been done.  The difficulty in extracting a scaling function
from data can be related to the phase of an orbit.  Unless
the parameter values are chosen with very large precision, the
experimental orbit will follow the theoretical one for a certain
period, deviate from it for short while, and once again follow
it for another period, repeating the cycle.  It appears that the
experimental orbit and the theoretical one are ``out of
phase''.

By using the golden mean mode locking as a concrete example, I
will show how the out of phase problem originates and how it can
be remedied so that the scaling function can be extracted from the
experimental data.  The method of solution consists of utilizing
data sets from different regions in parameter space to
reconstruct the dynamical system and its parameter dependence. 
The reconstructed dynamical system is then used to obtain the
scaling function.  In this process, care must be taken as to not
extract from the reconstructed system more information than what
was available from the data.

\begin{figure}
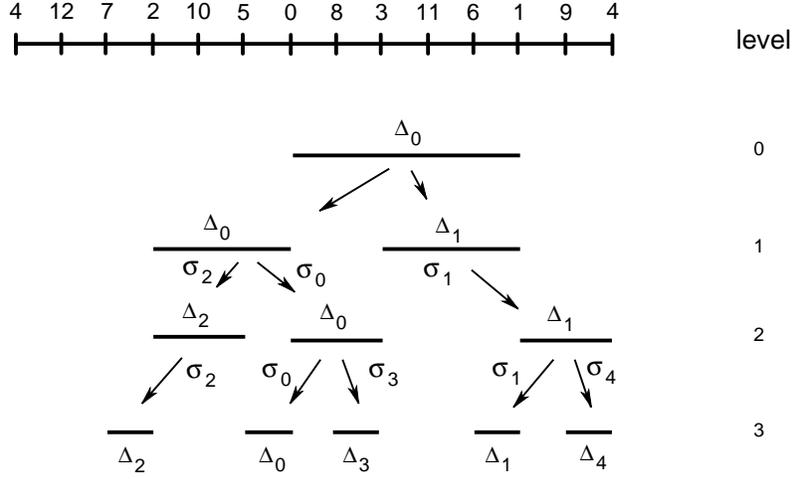

	\pfig{file=segments.ai,height=2.5in}
	\caption[boo]{Arrangement of the segments $\Delta^{(n)}_k$ used for
	constructing the scaling function.  The levels $n$ are
	indicated to the right side.  The scaling function is computed
	from the ratio of segments connected by arrows, the quantities
	$\sigma^{(n)}_k$.}
	\label{FigSegments}
\end{figure}

To understand the difficulties in extracting the scaling function we
must first understand how it is defined and how it is extracted from
data.  The basic ingredient in the definition is the partition of the
configuration space into segments.  To construct it for the circle map,
one has to consider the orbit of the inflection point.  As the
inflection point orbit rotates around the circle it delimits a series
of segments $\{ \Delta_k \}$.  If the rotation number is rational, the
segments will form a partition with a finite number of segments, as the
orbit is finite and every orbit point maps into another.  If the
rotation number is irrational, then one has to construct approximate
partitions by considering periodic orbits that resemble the orbit with
irrational rotation number.  The most effective way to approximate an
irrational number $\rho$ by fractions is to consider its continued
fraction expansion $\rho = [ a_1, a_2, a_3, \ldots ]$ and truncate it
after $n$ entries $P_n/Q_n = [ a_1, \ldots , a_n ]$.  One then proceeds
with the partition $\{\Delta_k^{(n)}\}$ of the circle for a map with
irrational rotation number $\rho$, as if the rotation number where
$P_n/Q_n$.  There will be a small error made, as the orbit is not
really periodic, but as observed by Shenker \cite{ShenkerScale}, the
error decreases exponentially fast at a universal rate of
$\alpha^{-n}$, with $\alpha = 1.28857$  An example of this partition is
shown in figure \ref{FigSegments}.  There the rotation number is the
golden mean $\rho_g = [1,1,1,\ldots ]$, with truncations given by
ratios of consecutive Fibonacci numbers, $1/2, 2/3, 3/5, \ldots$

The scaling function is computed from ratios of the segments in
$\{\Delta^{(n+1)}_t\}$ with those in $\{\Delta^{(n)}_t\}$.  One does
not define the scaling function directly.  First its value at different
points is defined, and then these are used to approximate the
function.  The preferred form of approximation is through piecewise
constant steps of height $\sigma^{(n)}_t$ placed in ascending order of
the integer $t$ and rescaled to approximate a function of the unit
interval.  Each scaling $\sigma^{(n)}_t$ is given by the ratio between
the size of the segment $|\Delta_{t}^{(n+ 1)}|$, and its parent
segment:
\bee
        \sigma^{(n)}_t =
        \frac
           {\displaystyle | \Delta_{t}^{(n+1)} | }
           {\displaystyle | \Delta^{(n)}_{\Theta(t,F_{n})} | }
        \mcomma
\eee
where $F_{n}$ is the Fibonacci number of segments that are used at
level $n$.  The function $\Theta(t,F)$ is the parent index function
which in the simple case of the golden-mean returns $t-F$ if $t\geq F$
and $t$ otherwise.  Formulas for other rotation numbers are given in
the appendix  of Ecke {\em et al.\/}~\cite{EckeRayleigh}.  This
definition differs from the one given by Feigenbaum \cite{FeigenbaumCMapPresentation} only by the
use of forward iterates of the map rather than backwards iterates
as he does.

Given a time series, a Poincar\'{e} section of the phase space
of the dynamical system is reconstructed using time delay
coordinates, and in the case of circle maps, the orbit lays on
a manifold that should be diffeomorphic to a circle --- a loop.  To
determine the scaling function, an orbit with rotation number
ending in a series of ones must be found.  For the fastest
convergence the orbit with rotation number $[1,1,1,1,\ldots]$ is
preferred, but in many cases it is outside the experimentally
accessible region in parameter space.  In this case any rotation
number ending in a series of ones is suitable, as the universal
numbers are asymptotical approached as the number of ones in the
continued fraction tail goes to infinity.  The smaller the
initial sequence of numbers different from one, the faster the convergence to the
universal values (if any).

For the determination of the scaling function the rotation number must
be known precisely, which is never the case in an experiment where the
orbits are always periodic within some experimental error.  To
illustrate the difficulties that this may cause, let us consider the
orbit of the sine circle-map, $\theta_{n+1} = \theta_{n }+ \Omega -
\sin(2\pi\theta_{n})/(2\pi)$ (notice that the constant that usually
multiplies the sine function has been taken to be $1$).  The parameter
$\Omega$ is chosen so that the rotation number is $76/351 =
[1,4,1^7,3]$, where $1^7$ means the one repeated seven times.  This
orbit occurs at the parameter value $\Omega=0.258971$ of the sine
circle map.  There are two other rotation numbers related to $76/351$
that are relevant: a closest golden mean tail irrational and the
closest fraction that is a golden mean approximant.  A golden mean tail
irrational is a continued fraction with the same initial sequence and
terminating in an infinite series of 1{\em s}, which for $76/351$ would
be the irrational $[1,4,1,1,1,\ldots]= (7-\sqrt{5})/22$; it occurs at
parameter value $\Omega=0.258978$.  The closest fraction is obtained by
keeping the longest sequence of 1{\em s} in the continued fraction
expansion of the rotation number $76/351$, and is the fraction $21/97 =
[1,4,1^7]$; it occurs at parameter value $\Omega=0.258956$.  The
parameter values for all these rotation numbers are close by, differing
by at most $1.5 \times 10^{-4}$.  With this small difference the
$76/351$ could be confused with the irrational orbit within
experimental errors.

\begin{figure}
	\pfig{file=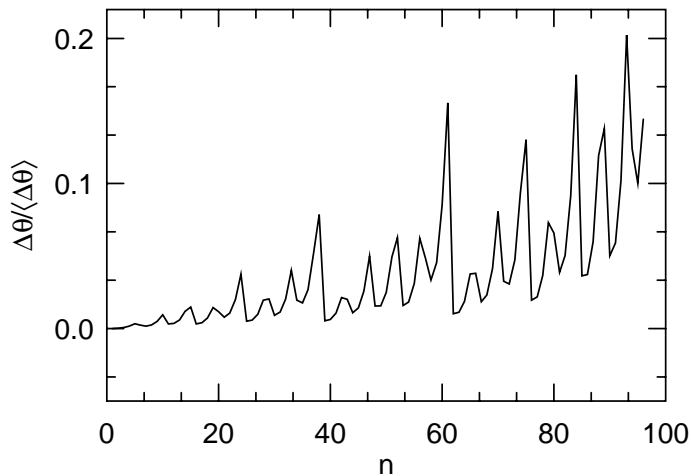,height=2.5in}
	\caption{Normalized error between actual orbit and approximate one.}
	\label{PhasePlot}
\end{figure}

Suppose that the $76/351$ orbit is mistaken for the orbit with
irrational winding number.  This means that one would construct the
scaling function using one of the approximants of the irrational, in
this case the fraction $21/97$.  One would consider the first 97 points
of the the orbit and try and extract the scaling function from it.  To
make maximum use of the data, and to compensate for the distortion of
the circle in the reconstructed data, the extra points could be use to
determine the arc length along the loop.  This techniques fails because
the first 97 points of the longer orbit match only the first twenty or
so points of the orbit at the irrational winding number.  As the orbit
proceeds the experimental $76/351$ orbit, $x_n^{\rm e}$, and the actual
golden mean orbit, $x_n^{\rm a}$, get out of phase.  This is
illustrated in figure \ref{PhasePlot} where the difference between the
golden mean tail orbit and the approximate orbit (the
76/351 cycle) is plotted as a function of the number of iterations. The
difference $\Delta x_n = |x_n^{\rm e} - x_n^{\rm a}|$ has been
normalized by the average separation between neighboring points
$\langle \Delta x \rangle$, the relevant scale in computing the scaling
function.  From the plot it is clear that any attempt to obtain the
scaling function directly from the 76/351 orbit will fail.

To circumvent the phase problem one can use
the fact that that several data sets with nearby parameters are
available and use these to reconstruct not only the return map at
a given parameter value, but also the dependence of the map on
the control parameters.  This will allow us to interpolate the
the map for the exact control parameter and then use the
interpolated map to obtain the scaling function.

The return map is constructed by using arc length along the loop and
determining the action of the map on it; the arc length is normalized
to one.  This return map then gives us a series of points that belong
to a function of the unit interval to itself.  To be able to iterate
the map the function must be known on the whole unit interval, so one
must choose a class of functions and find a representative of that
class that has as members the pairs of data points.  The function must
be at least twice differentiable if it is to be able to approximate
maps of the universality class of the sine circle map, it must be
periodic in its argument, and it must be monotonic.  Several classes of
functions that at first seem appropriate must be rejected.  Polynomials
cannot be used because in trying to reproduce the data a large degree
polynomial must be used and there is no direct way to assure that it will
be monotonic.  Splines will also tend not to be monotonic.  There is a
class of splines that are monotonic, but they usually lead to a system
of nonlinear equations which are difficult to solve.  

To solve these problems  determine a function $g$ that comes
sufficiently close to the data points and minimizes the wiggling. The
function is chosen from the class of cubic splines with knots
(suggested endpoints for the splines) at the data abscissae (for a
practical introduction to splines, see the book by de Boor
\cite{deBoorPractical}).  For $g$ to approximate the data points I
require a least square condition
\bee
	\sum_i p_i(y_i-g(x_i))^2  < C
\eee
where the $p_i$ is the relative weight of the data point
$(x_i,y_i)$ and $C$ a maximum error.  To minimize the wiggling
I require that the integrated curvature square be minimized,
that is,
\bee
        \int_0^1 dx\, |g''(x)|^2
\eee
be a minimum.  I choose to minimize the curvature rather than the sum
of square of the errors because I wanted to able to control how close
the function $g$ came to the data set.  Also, it is simpler to minimize
the curvature with the least square condition as a constraint than to
do it the other way around.  The standard approach for solving this
minimization problem is to introduce a Lagrange multiplier $\lambda$
and solve the combined problem, but this leads to a non-linear problem
in $\lambda$.  This can be avoided if we notice that the weights $p_i$
are proportional to the third derivative of $g$ computed at the
abscissae.  This leads to a linear problem that can be solved with
sparse matrix techniques (relevant here because large data sets can
lead to large matrices in the spline problem). The curvature
minimization algorithm does not guarantee that the map obtained will be
monotonic, but I found in practice that if the spline is within
experimental error to the data, the resulting map is monotonic.

\begin{figure}
	\pfig{file=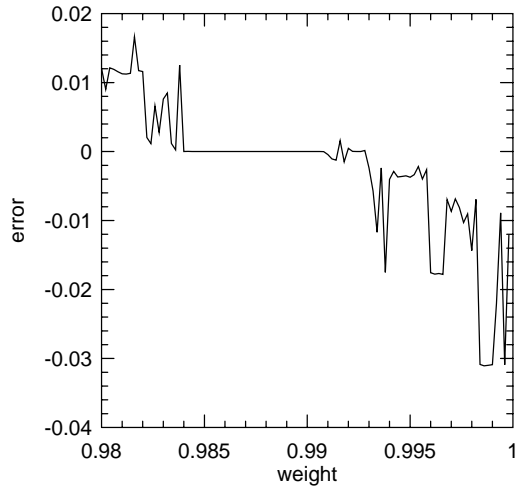,height=2.5in}
	\caption{Error in closing the $21/97$ orbit for the
	interpolated map as a function of the interpolating
	weight.  One can distinguish a region between 0.9846 and
	0.9906 where the map is mode-locked.}
	\label{FigError}
\end{figure}

Having determined two data sets that are close by in parameter space
(for the data set, the parameters cannot be resolved except for their
rotation number) one can proceed to determine an interpolated map that
has the exact rotation number.  Because the two maps that are used for
interpolation are so close together I use a straight forward linear
interpolation between the ordinates of the two maps.  If the ordinates
of one of the maps are represented by $y_n$ and the ordinates of the
other are represented by $z_n$, then the ordinate of the interpolated
map depends on a weight $c$ which varies between $0$ and $1$ and is
given by $c y_n + (1-c)z_n$.  Each interpolated map is a piecewise
cubic polynomial and is iterated to determine its rotation number.  If
the error in a given cycle not closing is plotted as a function of the
interpolation parameter, one can clearly distinguish the mode locked
region.  In figure \ref{FigError} the experimental maps for a $21/97$
cycle and a $34/129$ cycle are linearly interpolated for several
weights (with $c=1$ being the pure $21/97$ cycle).  For each
interpolated map the error in a $97$ cycle closing is plotted (the
distance between points $x_{600}$ and $x_{600+97}$).  The mode locked
region can be seen in the plot as the region where the error goes to
zero.  From within the mode locked region one must choose the parameter
value where the inflection point of the map is part of the orbit ---
the superstable parameter point.  The use of cubic splines precludes
using the second derivative of the map to determine the superstable
point, as the derivative is not very smooth.  I settled for using the
middle of the tongue as the superstable point, which leads to
acceptable results.

From the superstable interpolated map the scaling function can be
computed.  With the interpolation method I can determine periods
of lengths limited only by the computer capabilities, but
I have been careful not to use periods that lead to average
segment sizes $\langle \Delta_k \rangle$ that are smaller
than the original segments.  If this precaution is not taken
the method will generate orbits whose universality class is dictated by
the nature of the spline, rather than the data.  

\begin{figure}
	\pfig{file=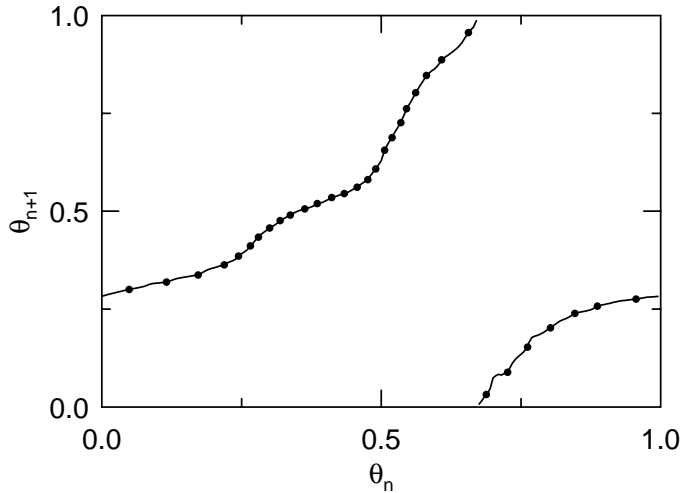,height=2.5in}
	\caption{Splined return map (solid) and original data (dots).}
	\label{SplinedMap}
\end{figure}

\begin{figure}
	\pfig{file=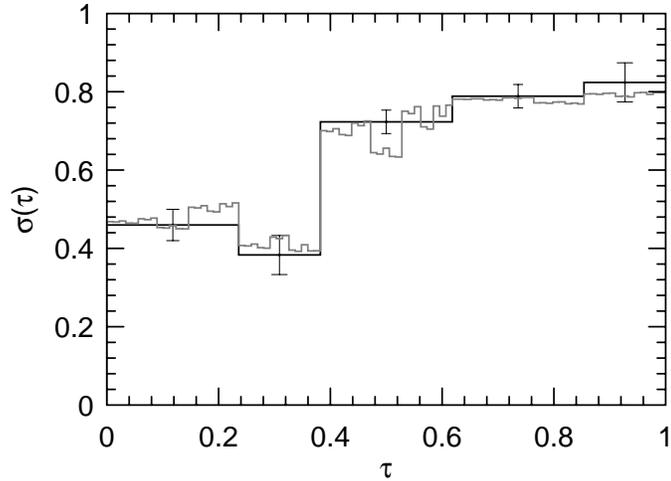,height=2.5in}
	\caption{Experimental (in gray) and theoretical (in black) scaling
	functions.  The large steps of the experimental scaling
	function should be compared at the left part of each step.}
	\label{ScalingFunction}
\end{figure}

The result of the reconstruction of the return map is shown in figure
\ref{SplinedMap}.  The original data set is shown as dots and the
resulting smoothed splined map is shown as a solid curve.  The two
different data sets that where used to reconstruct the return map are
not plotted, as on the scale of the plot they are not distinguishable
from the final map.
The scaling function resulting from reconstruction of the data is
shown in figure \ref{ScalingFunction}.  For comparison the
theoretical scaling function for the universality class of the
sine circle map is shown in the same plot. Notice
that scaling function plotted is not the result of averaging over
several data sets, but just of one long orbit.  The
error bars are estimated based on several different rotation
numbers with golden mean tail.  

I would like to conclude that even though trajectory scaling functions
are complicated to define (specially when compared to the $f(\alpha)$
spectrum of scalings) it is possible to extract them from experimental
data.  To be successful one must be careful that the control parameters
are tuned with enough precision for the length of the orbit used.

This work was funded by the Department of Energy.

\end{document}